\documentclass[12pt]{iopart}

\expandafter\let\csname equation*\endcsname=\relax
\expandafter\let\csname endequation*\endcsname=\relax

\usepackage{amsmath}

\usepackage{amssymb}

\usepackage{graphicx}

\usepackage{dcolumn}

\usepackage{bm}

\usepackage{xcolor}

\usepackage[export]{adjustbox}

\usepackage[T1]{fontenc}

\newcommand{\EqRef}[1]{(\ref{#1})}

\newcommand{\energy}{E}

\newcommand{\GkEsigma}  {  {\cal{G}}^{\sigma}     _{\energy  \bm{k}}  }
\newcommand{\GkEup}     {  {\cal{G}}^{\uparrow}   _{\energy  \bm{k}}  }
\newcommand{\GkEdown}   {  {\cal{G}}^{\downarrow} _{\energy  \bm{k}}  }

\newcommand{\GkprimeEup}     {  {\cal{G}}^{\uparrow}   _{\energy  \bm{k}'}  }
\newcommand{\GkprimeEdown}   {  {\cal{G}}^{\downarrow} _{\energy  \bm{k}'}  }

\newcommand{\Integral}{ \int_{-\infty}^{\; \energy_{F}} d\energy \,}

\newcommand{\TraceL}[1]
{
  \mathrm{Tr_{L}} \,
  \Bigg\{
    #1
  \Bigg\}
}

\newcommand{\PTerm}[1]
{
   \Big(
         \sum_{p} \,
           #1
   \Big)
}

\newcommand{\sublattice}[1]{\tilde{#1}}

\newcommand{\regexp}[1]
{
\mathrm{exp} 
   \big\{
     #1
   \big\}
}

\newcommand{\regcos}{ \mathrm{cos} }
\newcommand{\regsin}{ \mathrm{sin} }

\newcommand{\eqdot}{\,\,\, .}

\newcommand{\eqcomma}{\,\,\, ,}

\newcommand{\FGT}{Fe$_3$GeTe$_2$~}

\begin{document}
       


\title[]
{
\textit{Ab initio} study of anisotropic effects in two-dimensional \FGT using $\bm{k}$-dependent Green's functions
}



\author{Ilya~V.~Kashin}

\address
{
Theoretical Physics and Applied Mathematics Department, Ural Federal University, Mira Str. 19, 620002 Ekaterinburg, Russia
}

\ead{i.v.kashin@urfu.ru~(corresponding author)}


\author{Sergei~N.~Andreev}

\address
{
Theoretical Physics and Applied Mathematics Department, Ural Federal University, Mira Str. 19, 620002 Ekaterinburg, Russia
}

\ead{s.n.andreev@urfu.ru}


\vspace{10pt}
\begin{indented}
\item[]February 2025
\end{indented}



\begin{abstract}
In the present work, we develop the Green's function apparatus and extend its applicability to the study of microscopic anisotropic effects in real conducting materials. 
The problem of the previously proposed approaches written in terms of inter-atomic Green's functions is the presence of a spatial sum over all atoms of the crystal, which greatly complicates their application to systems with itinerant electrons. 
To provide a solution we derived expressions for magnetic torque vector and Dzyaloshinskii-Moriya interactions based on $\bm{k}$-dependent Green's functions, which allow numerical evaluations with guaranteed stability of spatial sums over the crystal lattice and moreover with much lower computational cost. 
Approbation of the approaches on the case of \FGT monolayer, which is based on first-principles DFT calculations, confirmed the numerical stability and allowed us to reproduce the characteristic length of experimentally observed collective spin excitations in the domain structure of this promising conducting material.
\end{abstract}



\vspace{2pc}






\section{Introduction}

Today, magnetic materials in which topological collective magnetic excitations (commonly skyrmions) are experimentally observed attract undying attention of scientists. 
This interest is fuelled not only by the well-known prospects for the construction of spintronics devices 
\cite{10.1063/5.0072735} 
and high-density and energy-efficient computer memory modules 
\cite{10.1126/science.1240573, 10.1063/5.0046950}. 
Recently, skyrmions have found their application in the task of noise suppression during quantum computation, which makes the problem of studying the mechanisms of such excitations even more relevant 
\cite{Ornelas2025}.

In this regard, two-dimensional materials consistently take a special place, following the experimental synthesis of van der Waals magnets, such as CrI$_3$ 
\cite{Huang_2017}
and 
Cr$_2$Ge$_2$Te$_6$
\cite{Gong_2017}.
Being highly sensitive to the electric field 
\cite{Jiang2018, Huang_2018}
and mechanical manipulations 
\cite{PhysRevB.98.144411}, 
these class of materials opens up new prospects towards building the ultra-compact spintronic devices 
\cite{adma.201900065}. 
\FGT monolayer is somehow outstanding representative of this class due to the metallic behavior, while for the most 2D magnets insulating or semiconducting electronic structure is commonly observed.
Accompanied by relatively high Curie temperature 
$T_C \sim 220$ K 
\cite{Deiseroth2006Fe3GeTe2AN, Deng_2018, Chen_2013} 
down to the monolayer limit 
\cite{Fei}, 
this feature set the magnetotransport properties especially valuable in spintronics 
\cite{Kim_2018, Roemer_2020}.

It is known that the crystal structure of the \FGT monolayer is characterized by non-equivalent Fe sublattices, whose chemical environment has a broken inversion symmetry and hence opens prospects of stabilizing the chiral magnetic structures 
\cite{PhysRevB.102.060402}. 
Despite the intrinsic planar symmetry of the mirror-related Fe sublattices, which basically suppresses this kind of collective spin excitations in view of the unit cell, the experimental studies report the presence of the magnetic domain structure and formation of the skyrmions 
\cite{acs.nanolett.9b03453, PhysRevB.103.104410, Birch, wang2019directobservationschiralspin}.
Therefore, the mechanisms, characteristics and origins of the magnetic structures are yet to be understood.

Our study is devoted to lifting the veil from this problem by employing first-principles DFT calculations, followed by engagement of Green's functions apparatus to provide numerical estimations for the basic characteristics of these structures.
Previously suggested approaches for magnetic torque vector (MTV)
\cite{PhysRevB.71.184434} 
and Dzyaloshinskii-Moriya interactions (DMI)
\cite{PhysRevB.89.214422} 
were expressed in terms of inter-atomic Green's functions and then imply pairwise aggregation of the contributions of each atom of the lattice to provide every single estimation. 
Such spatial sums over the atoms are known 
\cite{PhysRevB.64.174402, KASHIN201858, PhysRevB.106.134434} 
to suffer poor or even absent convergence if one deals with real conducting material. 
It could be established as the reason why these approaches appear not very popular among researchers, which is to say that there is no reliable convergence criterion provided by the approaches themselves.

To improve this situation in present study we derive new expressions for MTV and DMI on the base of Green's function with reciprocal space resolution. It was previously demonstrated that this transformation could result in almost complete solution of mentioned convergence problem during estimation of isotropic pairwise exchange interactions 
\cite{PhysRevB.106.134434} 
and magnetocrystalline anisotropy energy 
\cite{kashin2024magnetocrystalline}.
In this work we successfully expand this practice on the cases of MTV and DMI.
The approbation of the derived approaches allowed to originally reproduce the spatial length of the collective spin excitations in \FGT monolayer in accordance with the experimental data, which confirms new prospects of Green's functions apparatus to become useful instrument in investigations of the anisotropic effects in the real conducting materials.







\section{Method}

Our theoretical investigation should be started from the low-energy model of magnetoactive electron shell. 
In order to construct it one should project the wave functions, obtained from the first-principles calculations, into the basis of the Wannier functions. 
The result is commonly presented in the form of Hamiltonian, written in tight binding approximation as
\begin{equation}
\label{Eq:TBModel}
      \hat{H} =
            \sum_{i \ne j}
                  \sum_{\alpha \beta}
                        \sum_{\sigma}
                          \, t^{\sigma}_{i(\alpha) \, j(\beta)}
                                    \,
                              \hat{c}^{\dagger}_{i(\alpha) \sigma}
                                        \hat{c}_{j(\beta)  \sigma}
             + 
            \sum_{i}
                  \sum_{\alpha}
                        \sum_{\sigma}
                           \, \varepsilon^{\sigma}_{i(\alpha)}
                                    \,
                              \hat{c}^{\dagger}_{i(\alpha) \sigma}
                                        \hat{c}_{i(\alpha) \sigma} \eqcomma
\end{equation}
where 
$i$, $j$ denote atoms of the crystal;
$\alpha$, $\beta$ specify the orbital of the atom;
$\sigma = \uparrow, \downarrow$ identifies the spin projection of the electron;
$\hat{c}^{\dagger}_{i(\alpha) \sigma}$,
$\hat{c}_{j(\beta) \sigma}$ are creation and annihilation operators of the electron;
$\varepsilon^{\sigma}_{i(\alpha)}$ is the intra-atomic electron energy;
$t^{\sigma}_{i(\alpha) \, j(\beta)}$ is the hopping integral.

Once this Hamiltonian is represented in a matrix form, we can address it as how the unit cell with the translation vector $\bm{T} = 0$ interacts with the unit cell with arbitrary $\bm{T}$.
The size of this matrix is determined by amount of atoms in the unit cell and their orbital structure. Thus, the matrix sector $[H^{\sigma}( \bm{T} )]_{ij}$ describes the relation between atom $i$ (unit cell $\bm{T} = 0$) and atom $j$ (unit cell $\bm{T}$).

As the next step we transform this Hamiltonian to reciprocal space $H^{\sigma}(\bm{k})$, where
$\bm{k}$ denotes the reciprocal space vector of the Monkhorst-Pack grid 
\cite{PhysRevB.13.5188}.
The atomic structure of $H^{\sigma}(\bm{k})$ should be understood in terms of sublattices.
We call as sublattice the composition of atoms, which have the same local positions in the unit cells. Hence the each sublattice is essentially Bravais lattice and total amount of sublattices is just number of the atoms in the unit cell.
In our consideration we denote as $\sublattice{i}$ the sublattice, that contains the atom $i$, and the matrix sector $\big[ H^{\sigma}(\bm{k}) \big]_{\; \sublattice{i} \sublattice{j}}$ captures the interaction between corresponding sublattices in their integrity.

Then the Hamiltonian $H^{\sigma}(\bm{k})$ is used in the definition of the Green's function
\begin{equation}
      \GkEsigma = \big\{ \energy - H^{\sigma}(\bm{k}) \big\} ^{-1} \eqcomma
\end{equation}
where 
$\energy$ should be assumed as the sweep energy 
with small imaginary part $i 0^{+}$ multiplied by identity matrix.
In order to construct the inter-atomic Green's function between atoms $i$ and $j$ one can apply the transformation
\begin{equation}
\label{Eq:InterSiteGF}
  G^{\sigma}_{ij} = 
    \frac{1}{ N_{\bm{k}} }  \,
    \sum_{ \bm{k} }         \,
    [\GkEsigma]_{\sublattice{i} \sublattice{j}}
      \cdot
    \regexp
    {
      -i \bm{k} (\bm{T}_{j} - \bm{T}_{i})
    }
     \eqcomma
\end{equation}
where $\bm{T}_{i}(\bm{T}_{j})$ is the translation vector of the unit cell, containing the atom $i$($j$).
In further consideration it is important to note that 
$[\GkEsigma]_{\sublattice{i} \sublattice{j}}$ is determined as for the corresponding pair of sublattices and does not depend on $\bm{T}_{i}$ and $\bm{T}_{j}$.

The derivation of the expressions for magnetic torque vector and DMI vector we perform on the theoretical ground that the spin-orbit coupling (SOC) is treated on the magnetoactive atoms in framework of a second-order perturbation theory 
\cite{PhysRevB.52.13419, PhysRevB.89.214422, PhysRevB.39.865, PhysRevB.47.14932, Goringe_1997},
and the spin rotations are considered small (around the ferromagnetic collinear ground state) in order to validate Andersen's ''local force theorem'' 
\cite{LocalForceTheorem_1, LocalForceTheorem_2, Lichtenstein2013correl13}.




   \subsection{Magnetic torque vector}

In the formalism of inter-atomic Green's functions the components of magnetic torque vector could be expressed as 
\cite{PhysRevB.71.184434} 
\begin{equation}
\begin{split}
\label{Eq:InitialTorque}
    A^{x}_{i} = 
- \frac{1}{2 \pi} 
  \Integral
  \mathrm{Re} \,
  \TraceL
  {
     \PTerm
     {
               &   \Delta_{i}      
           G^{\downarrow}_{ip}     
      H^{so}_{\downarrow 
              \uparrow}            
           G^{\uparrow}_{pi}      
     }
                -
     \PTerm
     {
                   \Delta_{i}      
           G^{\uparrow}_{ip}       
      H^{so}_{\uparrow 
              \downarrow}          
           G^{\downarrow}_{pi}   
     }
  }
            \\
    A^{y}_{i} = 
- \frac{1}{2 \pi} 
  \Integral
  \mathrm{Im} \,
  \TraceL
  {
     \PTerm
     {
               &   \Delta_{i}
           G^{\downarrow}_{ip}
      H^{so}_{\downarrow 
              \uparrow}
           G^{\uparrow}_{pi}
     }
                +
     \PTerm
     {
                 \Delta_{i}
           G^{\uparrow}_{ip}
      H^{so}_{\uparrow 
              \downarrow}
           G^{\downarrow}_{pi}
     }
  }
            \\
    A^{z}_{i} = 
- \frac{1}{4 \pi} 
  \Integral
  \mathrm{Re} \,
  \TraceL
  {
     \PTerm
     {
             &   \Delta_{i}
           G^{\uparrow}_{ip}
    ( H^{so}_{\uparrow   \uparrow} - 
      H^{so}_{\downarrow \downarrow} )
           G^{\downarrow}_{pi}
     }
                 \; - \\
                  & -
     \PTerm
     {
                   \Delta_{i}
           G^{\downarrow}_{ip}
    ( H^{so}_{\uparrow   \uparrow} - 
      H^{so}_{\downarrow \downarrow} )
           G^{\uparrow}_{pi}
     }
  }
   \eqcomma
\end{split}
\end{equation}
where 
$\energy_{F}$ is the Fermi energy,
$\mathrm{Tr_{L}}$ is the trace over the orbital index,
$p$ denotes the atom of the crystal and the corresponding sum constitutes the spatial surrounding of the $i$ atom,
$\Delta_{i}$ is the intra-atomic spin splitting 
$[H^{\uparrow}  ( \bm{T} = 0 )]_{ii} - 
 [H^{\downarrow}( \bm{T} = 0 )]_{ii}$,
$H^{so}~=~\lambda \, \bm{{\cal{L}}} \bm{{\cal{S}}}$ is the SOC operator for $d$ shell 
with $\bm{{\cal{L}}}$ and $\bm{{\cal{S}}}$ as the orbital momentum and the spin of the $d$ shell, correspondingly, and $\lambda$ is the small parameter.

Let us focus on the first integrand term of $A^{x}_{i}$ and split the atomic sum over $p$ into composition of the translation vector sum $\bm{T}_{p}$ and the sum over sublattices $\sublattice{p}$. 
Taking into account the definition of the inter-atomic Green's function \EqRef{Eq:InterSiteGF} one can write
\begin{equation}
\begin{split}
\label{Eq:TorquePTerm}
  \sum_{p} \,
                \Delta_{i}
        G^{\downarrow}_{ip}
   H^{so}_{\downarrow 
           \uparrow}
        G^{\uparrow}_{pi}
           =
  \sum_{ \sublattice{p} }             \,
  \frac{1}{ N_{\bm{k}} N_{\bm{k}'} }  \,
& \sum_{ \bm{k} \bm{k}' }             \,
                        \Delta_{i}                     \;
        [\GkEdown]_{\sublattice{i} \sublattice{p}}     \;
            H^{so}_{\downarrow \uparrow}               \;
        [\GkprimeEup]_{\sublattice{p} \sublattice{i}}  \;\; \times \\
         &\;\; \times
  \sum_{ \bm{T}_{p} } \,
  \regexp
  {
     i (\bm{k}' - \bm{k}) (\bm{T}_{p} - \bm{T}_{i})
  }
   \eqdot
\end{split}
\end{equation}

Then we establish that
\begin{equation}
  \sum_{ \bm{T}_{p} } \,
  \regexp
  {
     i (\bm{k}' - \bm{k}) (\bm{T}_{p} - \bm{T}_{i})
  }
   =
  N_{\bm{k}}  \;  \delta (\bm{k}' - \bm{k})
\end{equation}
and rewrite \EqRef{Eq:TorquePTerm} as
\begin{equation}
\label{Eq:TorqueReducedPTerm}
  \sum_{p} \,
                \Delta_{i}
        G^{\downarrow}_{ip}
   H^{so}_{\downarrow 
           \uparrow}
        G^{\uparrow}_{pi}
           =
  \frac{1}{ N_{\bm{k}} }        \,
  \sum_{ \sublattice{p} }       \,
  \sum_{ \bm{k} }               \,
                        \Delta_{i}                     \;
        [\GkEdown]_{\sublattice{i} \sublattice{p}}     \;
            H^{so}_{\downarrow \uparrow}               \;
        [\GkEup]_{\sublattice{p} \sublattice{i}}  
   \eqdot
\end{equation}

By applying the similar transformations to all $\sum_{p}$ terms in \EqRef{Eq:InitialTorque} we obtain new expressions for the components of the magnetic torque vector:
\begin{equation}
\begin{split}
\label{Eq:TorqueFinal}
    A^{x}_{i} = 
- \frac{1}{2 \pi} 
  \Integral
  \mathrm{Re} \,
  \TraceL
  {
      \frac{1}{ N_{\bm{k}} }        \,
      \sum_{ \sublattice{p} }       \,
      \sum_{ \bm{k} }               \,
                        \Delta_{i}                   \;   
        [\GkEdown]_{\sublattice{i} \sublattice{p}} & \;   
            H^{so}_{\downarrow \uparrow}             \;  
        [\GkEup]_{\sublattice{p} \sublattice{i}} 
                              -
                   \Delta_{i}                        \;   
        [\GkEup]_{\sublattice{i} \sublattice{p}}     \;   
            H^{so}_{\uparrow \downarrow}             \;   
        [\GkEdown]_{\sublattice{p} \sublattice{i}}
  }
            \\
    A^{y}_{i} = 
- \frac{1}{2 \pi} 
  \Integral
  \mathrm{Im} \,
  \TraceL
  {
      \frac{1}{ N_{\bm{k}} }        \,
      \sum_{ \sublattice{p} }       \,
      \sum_{ \bm{k} }               \,
                        \Delta_{i}                     \; 
        [\GkEdown]_{\sublattice{i} \sublattice{p}}  &  \;
            H^{so}_{\downarrow \uparrow}               \;
        [\GkEup]_{\sublattice{p} \sublattice{i}} 
                              +
                   \Delta_{i}                          \;
        [\GkEup]_{\sublattice{i} \sublattice{p}}       \; 
            H^{so}_{\uparrow \downarrow}               \; 
        [\GkEdown]_{\sublattice{p} \sublattice{i}}
  }
            \\
    A^{z}_{i} = 
- \frac{1}{4 \pi} 
  \Integral
  \mathrm{Re} \,
  \TraceL
  {
      \frac{1}{ N_{\bm{k}} }        \,
      \sum_{ \sublattice{p} }       \,
      \sum_{ \bm{k} }               \,
                 \Delta_{i}                           \;
        [\GkEup]_{\sublattice{i} \sublattice{p}}  &   \;
    ( H^{so}_{\uparrow   \uparrow} -                   
      H^{so}_{\downarrow \downarrow} )                \;
        [\GkEdown]_{\sublattice{p} \sublattice{i}}
                \;  - \\
                    - \;
             &   \Delta_{i}                           \;
        [\GkEdown]_{\sublattice{i} \sublattice{p}}    \;
    ( H^{so}_{\uparrow   \uparrow} -                 
      H^{so}_{\downarrow \downarrow} )                \;
        [\GkEup]_{\sublattice{p} \sublattice{i}}
  }
   \eqdot
\end{split}
\end{equation}
One can readily see that the spatial sum over $p$ atoms, which contains hundreds of terms in case of the real metallic systems, is now replaced by the sum over sublattices $\sublattice{p}$ with the amount of terms equal to the number of atoms in a single unit cell. Therefore we can state thus derived approach to have a computational performance hundreds times faster than \EqRef{Eq:InitialTorque}. Moreover, the sum over $\sublattice{p}$ is self-possessed guaranteed numerical convergence, whereas $\sum_{p}$ in \EqRef{Eq:InitialTorque} is expected to lose this convergence if the metallic system is under consideration.





   \subsection{Dzyaloshinskii–Moriya interactions}

In order to represent all three components of DMI vector we firstly define the rotation of the SOC operator $H^{so}$ 
from $(0, 0, 1)$ 
to 
$(\sin{\theta} \cos{\varphi}, \,
  \sin{\theta} \sin{\varphi}, \,
  \cos{\theta} )$ 
as
\begin{equation}
\label{Eq:HsoRotation}
    H^{so}(\theta, \varphi) = U^{-1}(\theta, \varphi) \, H^{so} \, U(\theta, \varphi)   \eqcomma
\end{equation}
where
\begin{equation}
    U(\theta, \varphi) = 
    \begin{pmatrix}
    \phantom{-} \regcos{(\theta / 2)} \phantom{ \cdot e^{i \varphi} } &
                \regsin{(\theta / 2)} \cdot e^{-i \varphi}              \\ 
      - \regsin{(\theta / 2)} \cdot e^{i \varphi} & 
        \regcos{(\theta / 2)} \phantom{ \cdot e^{-i \varphi} }
    \end{pmatrix}
\end{equation}
is Wigner's rotation matrix, and
$U^{-1}(\theta, \varphi)$ is its inverse variant.

Starting from the formalism of inter-atomic Green's functions, the components of DMI vector are thereby could be found 
\cite{PhysRevB.89.214422}
as 
\begin{equation}
\begin{split}
\label{Eq:DMIinitial}
   D_{ij} (\theta, \varphi) & = 
- \frac{1}{8 \pi} 
  \Integral
  \mathrm{Re} \,
     \\
  \TraceL
  {
    &\PTerm
     {
                  \Delta_{j}
            G^{\uparrow}_{ji} 
                   \Delta_{i}   
                   \cdot
           G^{\downarrow}_{ip}                    \;   
      H^{so}_{\downarrow 
              \downarrow} (\theta, \varphi)       \;   
           G^{\downarrow}_{pj}      
     }
                    -
     \PTerm
     {
                  \Delta_{j}
          G^{\downarrow}_{ji} 
                   \Delta_{i}   
                   \cdot
           G^{\uparrow}_{ip}                    \;   
      H^{so}_{\uparrow 
              \uparrow} (\theta, \varphi)       \;   
           G^{\uparrow}_{pj}      
     }
                \;  + \\ +
    &\PTerm
     {
                  \Delta_{i}
          G^{\downarrow}_{ij} 
                   \Delta_{j}   
                   \cdot
           G^{\uparrow}_{jp}                    \;   
      H^{so}_{\uparrow 
              \uparrow} (\theta, \varphi)       \;   
           G^{\uparrow}_{pi}      
     }
                    -
     \PTerm
     {
                  \Delta_{i}
            G^{\uparrow}_{ij} 
                   \Delta_{j}   
                   \cdot
           G^{\downarrow}_{jp}                    \;   
      H^{so}_{\downarrow 
              \downarrow} (\theta, \varphi)       \;   
           G^{\downarrow}_{pi}      
     }
  }
\end{split}
\end{equation}
setting 
$(\theta, \varphi) = (0, 0)$ for $D^{z}_{ij}$, 
$( \pi/2,  0 )$              for $D^{x}_{ij}$ and 
$( \pi/2,  \pi/2 )$          for $D^{y}_{ij}$.

In a manner of \EqRef{Eq:TorquePTerm} we can write for each of the $\sum_{p}$ term
\begin{equation}
\begin{split}
\label{Eq:DMIPTerm}
  \sum_{p} \,
                  \Delta_{j}
            G^{\uparrow}_{ji} 
                   \Delta_{i}   
                 & \cdot
           G^{\downarrow}_{ip}                    \;   
      H^{so}_{\downarrow 
              \downarrow} (\theta, \varphi)       \;   
           G^{\downarrow}_{pj} 
           =
                  \Delta_{j}
            G^{\uparrow}_{ji} 
                   \Delta_{i}   
                 \; \cdot \\ \cdot
& \sum_{ \sublattice{p} }             \,
  \frac{1}{ N_{\bm{k}} N_{\bm{k}'} }  \,
  \sum_{ \bm{k} \bm{k}' }             \,
                        \Delta_{i}                     \;
        [\GkEdown]_{\sublattice{i} \sublattice{p}}     \;
            H^{so}_{\downarrow \downarrow}             \;
        [\GkprimeEdown]_{\sublattice{p} \sublattice{i}}  \;\; \times \\
         &\;\; \times
  \sum_{ \bm{T}_{p} } \,
  \regexp
  {
   - i \bm{k}   (\bm{T}_{p} - \bm{T}_{i})
   - i \bm{k}'  (\bm{T}_{j} - \bm{T}_{p})
  }
\end{split}
\end{equation}
and then state 
\begin{equation}
  \sum_{ \bm{T}_{p} } \,
  \regexp
  {
   - i \bm{k}   (\bm{T}_{p} - \bm{T}_{i})
   - i \bm{k}'  (\bm{T}_{j} - \bm{T}_{p})
  }
   =
  N_{\bm{k}}  \;  \delta (\bm{k}' - \bm{k}) 
    \cdot
    \regexp
    {
      -i \bm{k} (\bm{T}_{j} - \bm{T}_{i})
    }
    \; ,
\end{equation}
which finally leads us to the expression for DMI vector component
\begin{equation}
\begin{split}
\label{Eq:DMIfinal}
   D_{ij} (\theta, \varphi) & = 
- \frac{1}{8 \pi} 
  \Integral
  \mathrm{Re} \,
  \TraceL
  {
      \frac{1}{ N_{\bm{k}} }        \,
      \sum_{ \sublattice{p} }       \,
      \sum_{ \bm{k} }               \; \times \\
     \times \;  & \Delta_{j}
            G^{\uparrow}_{ji} 
                   \Delta_{i}   
                   \cdot
           [\GkEdown]_{\sublattice{i} \sublattice{p}}                    \;   
      H^{so}_{\downarrow 
              \downarrow} (\theta, \varphi)       \;   
           [\GkEdown]_{\sublattice{p} \sublattice{j}}  
                   \cdot
    \regexp
    {
      -i \bm{k} (\bm{T}_{j} - \bm{T}_{i})
    }
    \; - \\
  - \;
               &  \Delta_{j}
          G^{\downarrow}_{ji} 
                   \Delta_{i}   
                   \cdot
           [\GkEup]_{\sublattice{i} \sublattice{p}}       \;   
      H^{so}_{\uparrow 
              \uparrow} (\theta, \varphi)                 \;   
           [\GkEup]_{\sublattice{p} \sublattice{j}} 
                   \cdot
    \regexp
    {
      -i \bm{k} (\bm{T}_{j} - \bm{T}_{i})
    }
    \; + \\
  + \;
               &  \Delta_{i}
          G^{\downarrow}_{ij} 
                   \Delta_{j}   
                   \cdot
           [\GkEup]_{\sublattice{j} \sublattice{p}}          \;   
      H^{so}_{\uparrow 
              \uparrow} (\theta, \varphi)                    \;   
           [\GkEup]_{\sublattice{p} \sublattice{i}} 
                   \cdot
    \regexp
    {
      -i \bm{k} (\bm{T}_{i} - \bm{T}_{j})
    }
    \; - \\
  - \;
               &  \Delta_{i}
            G^{\uparrow}_{ij} 
                   \Delta_{j}   
                   \cdot
           [\GkEdown]_{\sublattice{j} \sublattice{p}}             \;   
      H^{so}_{\downarrow 
              \downarrow} (\theta, \varphi)                       \;   
           [\GkEdown]_{\sublattice{p} \sublattice{i}}  
                   \cdot
    \regexp
    {
      -i \bm{k} (\bm{T}_{i} - \bm{T}_{j})
    }
  }
  \eqcomma
\end{split}
\end{equation}
which inherits the same advantages over \EqRef{Eq:DMIinitial} as the derived expression for magnetic torque vector \EqRef{Eq:TorqueFinal} possess in comparison with \EqRef{Eq:InitialTorque}.
Noteworthy that the problem of slow (or absent) convergence of the atomic spatial sum in \EqRef{Eq:DMIinitial} appears even more severe, due to necessity to compute this sum for each pair of atoms $ij$ individually.








\section{Results and Discussion}

To validate the developed methods on the example of the \FGT monolayer, we performed first-principles calculations within the framework of the GGA approach 
\cite{PhysRevLett.77.3865}.
The resulting wave functions were used to construct a low-energy model by projecting onto maximally localized Wannier functions basis 
\cite{PhysRevB.56.12847, PhysRevB.65.035109}.
The details are provided in ~\ref{AppendixA}.

\begin{figure}[htbp]
\begin{center}
\includegraphics[width=0.9\textwidth]{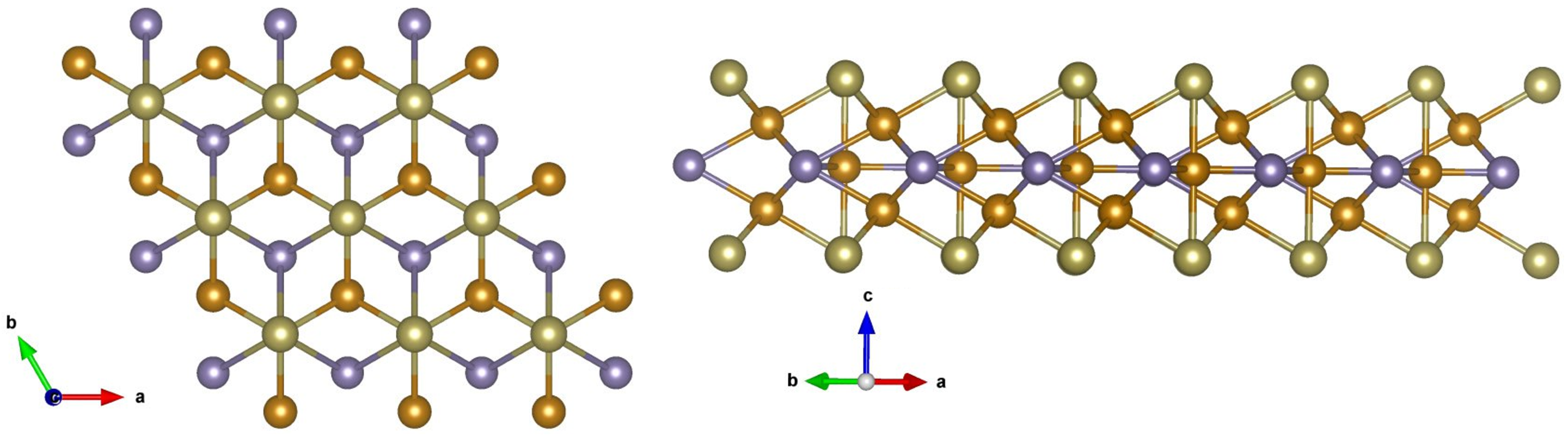}
\end{center}

\caption{\label{fig:FGTCrystalStructure} 
The crystal structure of \FGT monolayer.
Orange, purple and green spheres denote Fe, Ge, Te atoms, correspondingly.
Fe atoms constitute top, middle and bottom layer as three sublattices 
(local positions are presented in \ref{AppendixA}).
}
\end{figure}

\FGT monolayer crystallizes in the hexagonal structure with the point group $D_{6h}$ 
\cite{Deiseroth}. 
The key feature of the material is that the structure contains Fe atoms with different valence types. In figure 
\ref{fig:FGTCrystalStructure} 
Fe atoms constituting top and bottom layer are Fe$^{3+}$, while the middle layer contains Fe$^{2+}$ 
\cite{PUSHKAREV2023171456}.
On the picture of microscopic magnetism it is first of all manifested at the level of local magnetic moments, which are estimated as 
$2.39~\mu_B$ for Fe$^{3+}$ and 
$1.55~\mu_B$ for Fe$^{2+}$ in a good agreement with previous studies 
\cite{PhysRevB.102.060402, PUSHKAREV2023171456}.

\begin{figure}[htbp]
\begin{center}
\includegraphics[width=0.47\textwidth]{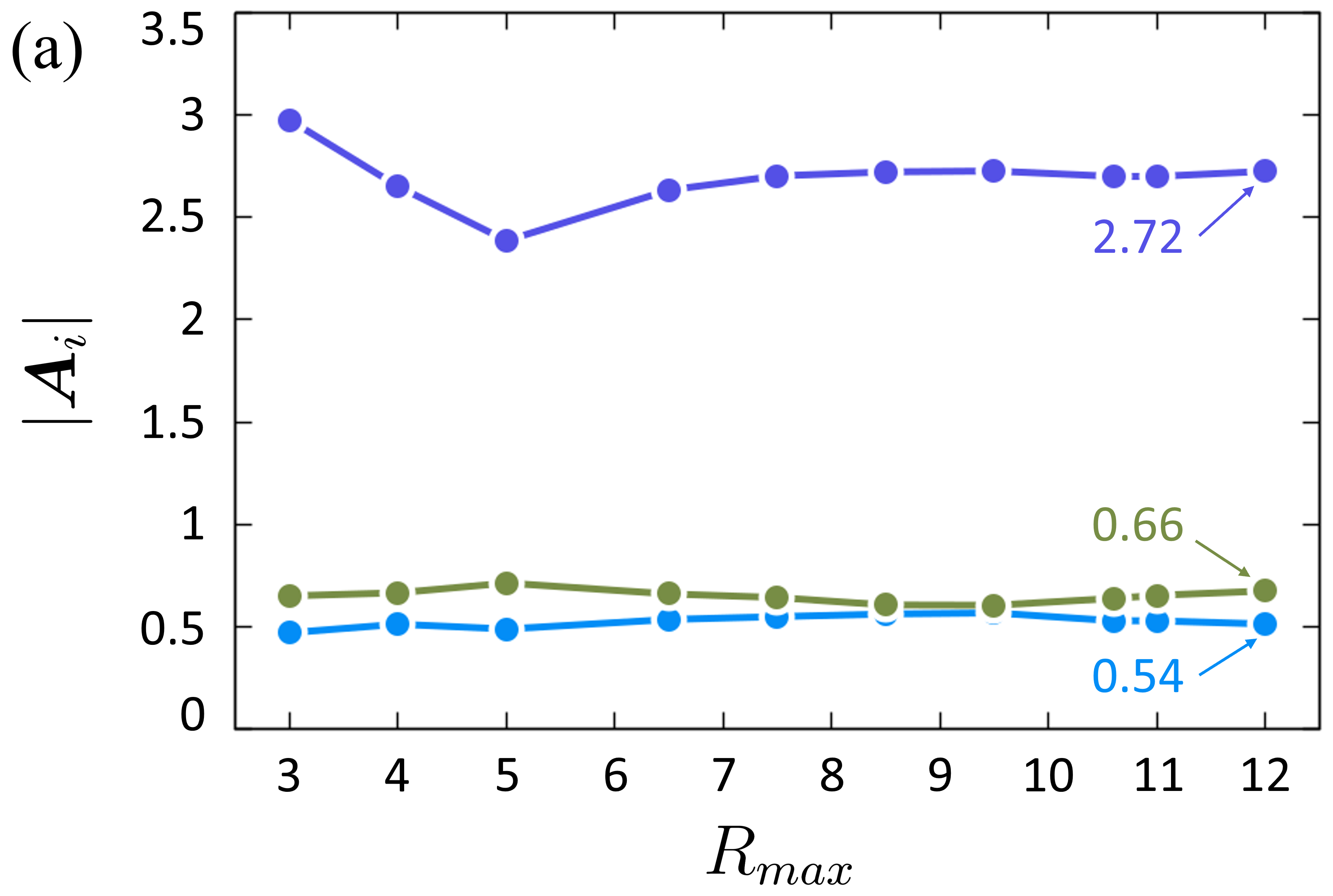}
\includegraphics[width=0.47\textwidth]{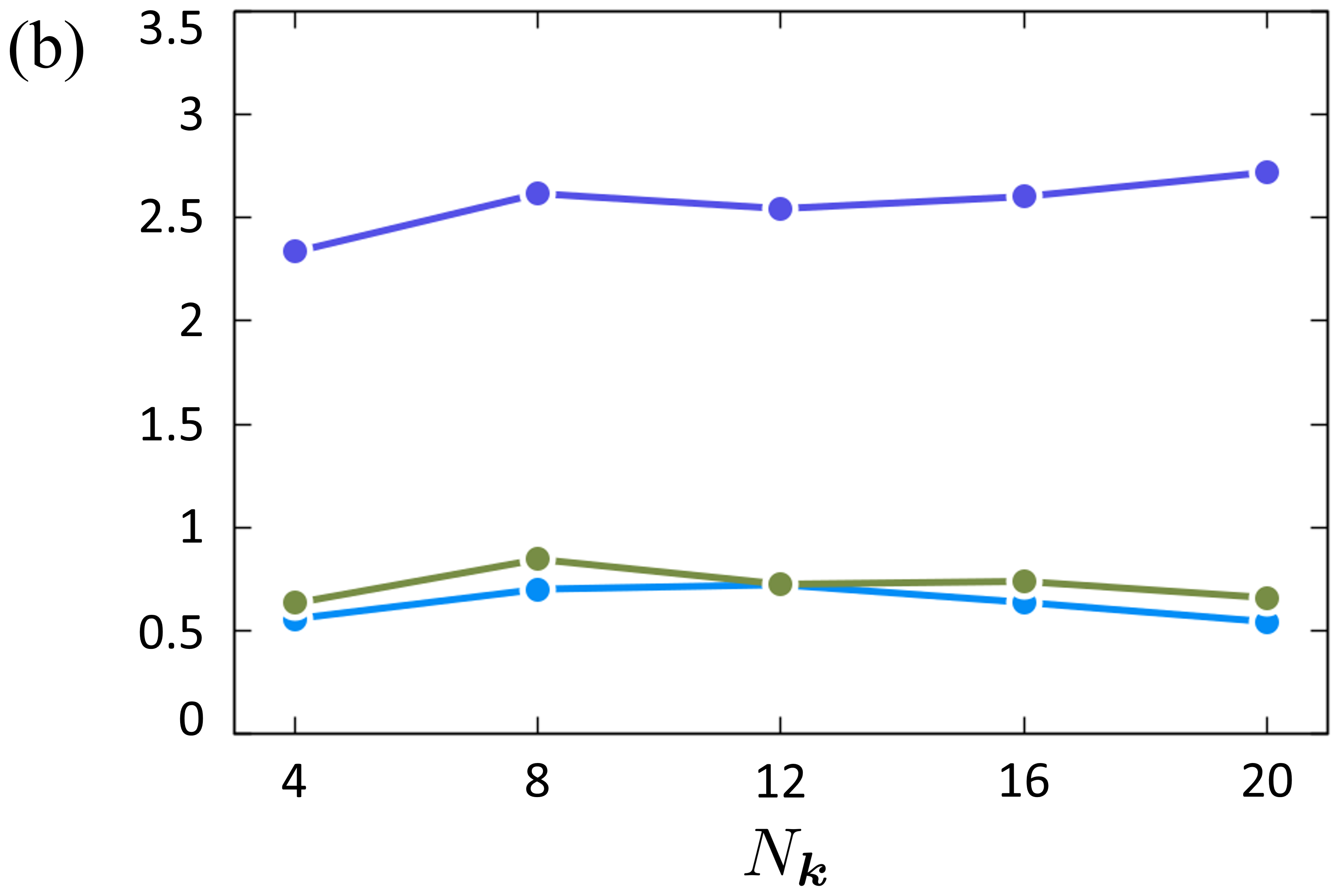}
\end{center}

\caption{\label{fig:FGTTorques} 
Obtained values of the magnetic torque vector $|\bm{A}_{i}|$ in \FGT monolayer, in meV.
Blue, green and purple color denote the atom Fe from top, bottom and middle layer, correspondingly 
(figure \ref{fig:FGTCrystalStructure}, local atomic positions are presented in \ref{AppendixA}).
(a)~Convergence dynamics of \EqRef{Eq:InitialTorque} as the maximum distance threshold ($R_{max}$, in \AA) of the sum $\sum_{p}$ grows. 
The arrowed values are the estimation using suggested approach \EqRef{Eq:TorqueFinal}.
(b)~Estimations performed using the suggested approach \EqRef{Eq:TorqueFinal} with Monkhorst-Pack grid density $N_{\bm{k}} \times N_{\bm{k}}$ (the arrowed values in (a) are those for the case of $20 \times 20$ grid).
}
\end{figure}

The estimated values of the magnetic torque vectors for three inequivalent Fe sublattices
\cite{PhysRevB.102.060402} 
are presented in 
figure~\ref{fig:FGTTorques}. 
Apart from the general stability by means of Monkhorst-Pack grid density 
(figure~\ref{fig:FGTTorques}(b)) 
we establish that 
the values obtained using the expression based on the inter-atomic Green's functions 
\EqRef{Eq:InitialTorque} 
are characterized by a slow convergence 
(figure~\ref{fig:FGTTorques}(a)) 
as the maximum distance threshold grows in corresponding cumulative sum $\sum_{p}$.
Meanwhile the converged result is directly obtained by using the suggested method 
\EqRef{Eq:TorqueFinal}.
Besides the obvious advantage of completely removing the problem of $\sum_{p}$ numerical convergence, our approach can serve as a useful complement for the original approach 
\EqRef{Eq:InitialTorque}, 
since it plays the role of a reliable convergence criterion.
It turns out to be especially important for conducting systems in which $\sum_{p}$ reaches convergence at distances of hundreds \AA~through passing local plateaus, which can be erroneously interpreted as the resulting values
\cite{PhysRevB.64.174402, KASHIN201858, PhysRevB.106.134434}.

In order to physically validate the obtained result we perform the theoretical estimation of the characteristic length of periodic collective spin excitations in \FGT monolayer. 
These excitations are experimentally known to be stripe domain structure observed using the x-ray
microscopy, and as its length we assume the double domain wall width with the value 80$-$240 nm 
\cite{Birch}.
For this purpose let us write an expression for the total energy variation around perfect ferromagnetic ground state 
\cite{PhysRevB.71.184434}:
\begin{equation}
\label{Eq:EnergyVariation}
    \Delta E =
         \sum_{i}
         \bigg[
            \bm{A}_{i} \cdot \bm{\delta \alpha}_{i}
                        +
            \frac{1}{2} \sum_{j \ne i} J_{ij} 
                \big|
                    \bm{\delta \alpha}_{i} - \bm{\delta \alpha}_{j}
                \big|
         \bigg]
         \eqcomma
\end{equation}
where 
$\sum_{i}$ is assumed to address the atoms in the unit cell with $\bm{T} = 0$,
$\sum_{j}$ is the spacial sum over all crystal's lattice,
$\bm{\delta \alpha}_{i}$($\bm{\delta \alpha}_{j}$) is the corresponding canting angle,
$J_{ij}$ is the isotropic exchange interaction.
Then we mention that the canting angle $\bm{\delta \alpha}_{i}$ generally characterizes the sublattice $\sublattice{i}$, while $J_{ij}$ is basically assumed to be the function of actual distance between atoms $i$ and $j$.

The straightforward way to estimate thus introduced spatial sum 
$\sum_{j \ne i} J_{ij}$ is to exhaust it term by term using infinitesimal spin rotations technique 
\cite{LIECHTENSTEIN198765}
\begin{equation}
\label{Eq:LKAG}
J_{ij} =
- \frac{1}{8 \pi} 
  \Integral
  \mathrm{Im} \,
  \TraceL
  {
    \sum_{ \sigma } \;
    \Delta_{i}  
    G^{\sigma}_{ij}
    \Delta_{j}  
    G^{- \sigma}_{ji}
  }
  \eqdot
\end{equation}
But this procedure faces the same problem of slow or even absent spatial convergence.
To circumvent it in 
\cite{PhysRevB.106.134434} the approach was proposed to consider these interactions in the reciprocal space:
\begin{equation}
[J(\bm{q})]_{\sublattice{i} \sublattice{j}} =
- \frac{1}{8 \pi} 
  \Integral
  \mathrm{Im} \,
  \TraceL
  {
      \sum_{ \sigma } \;
      \frac{1}{ N_{\bm{k}} }        \,
      \sum_{ \bm{k} }               \,
      \Delta_{i} \;
     [{\cal{G}}^{\sigma}_{\energy  \;  \bm{k + q}}]_{\sublattice{i} \sublattice{j}}   \;
      \Delta_{j} \;
     [{\cal{G}}^{- \sigma}_{\energy  \bm{k}}]_{\sublattice{j} \sublattice{i}}   \;
  }
  \eqcomma
\end{equation}
where $\bm{q}$ is the corresponding reciprocal space vector.
Taking into account that $J_{ij}$ and $J(\bm{q})$ are interrelated by means of Fourier transform, we can write the expression 
\begin{equation}
\sum_{j \ne i} J_{ij} =
    \sum_{ \sublattice{j} } \; 
    [J(\bm{q} = 0)]_{\sublattice{i} \sublattice{j}}
            -
     J_{ii}
     \eqcomma
\end{equation}
where intra-atomic parameter $J_{ii}$ can be found using \EqRef{Eq:LKAG}.
For our case we finally obtain 
$\sum_{j \ne 1} J_{1j}$ = 71.4~meV,
$\sum_{j \ne 2} J_{2j}$ = 64.9~meV,
$\sum_{j \ne 3} J_{3j}$ = 40.3~meV.
The mean-field assessment 
\cite{PhysRevB.64.174402} 
of Curie temperature 
is 312~K, which is in a good agreement with the previous theoretical studies 
\cite{Badrtdinov}, but overestimates the experimental value 200~K 
\cite{Deng, Zaiyao}.
Nevertheless, the minimization of 
\EqRef{Eq:EnergyVariation} using generalized reduced gradient method and further estimation of the periodic spin excitations length as 
$2 \pi L / |\bm{\delta \alpha}_{i}|$ 
($L$~=~0.399~nm is the distance between the nearest neighbors in the sublattices)
yields 
116, 190 and 61~nm~for three Fe sublattices, which agrees well with the experimental measurements and thus basically confirms the suggested method applicable to the real conducting material such as \FGT monolayer.

\begin{figure}[htbp]
\begin{center}
\includegraphics[width=0.47\textwidth]{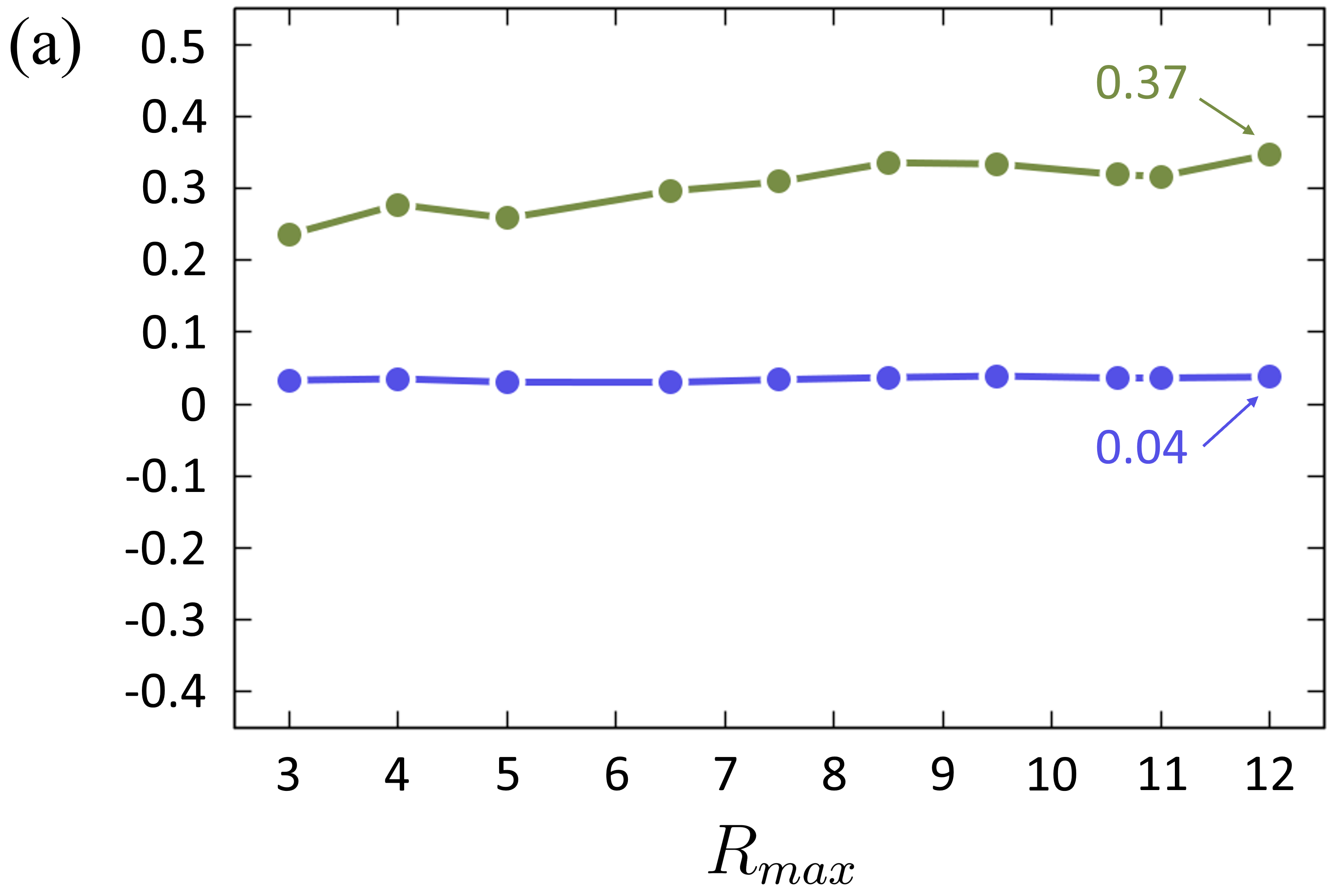}
\includegraphics[width=0.47\textwidth]{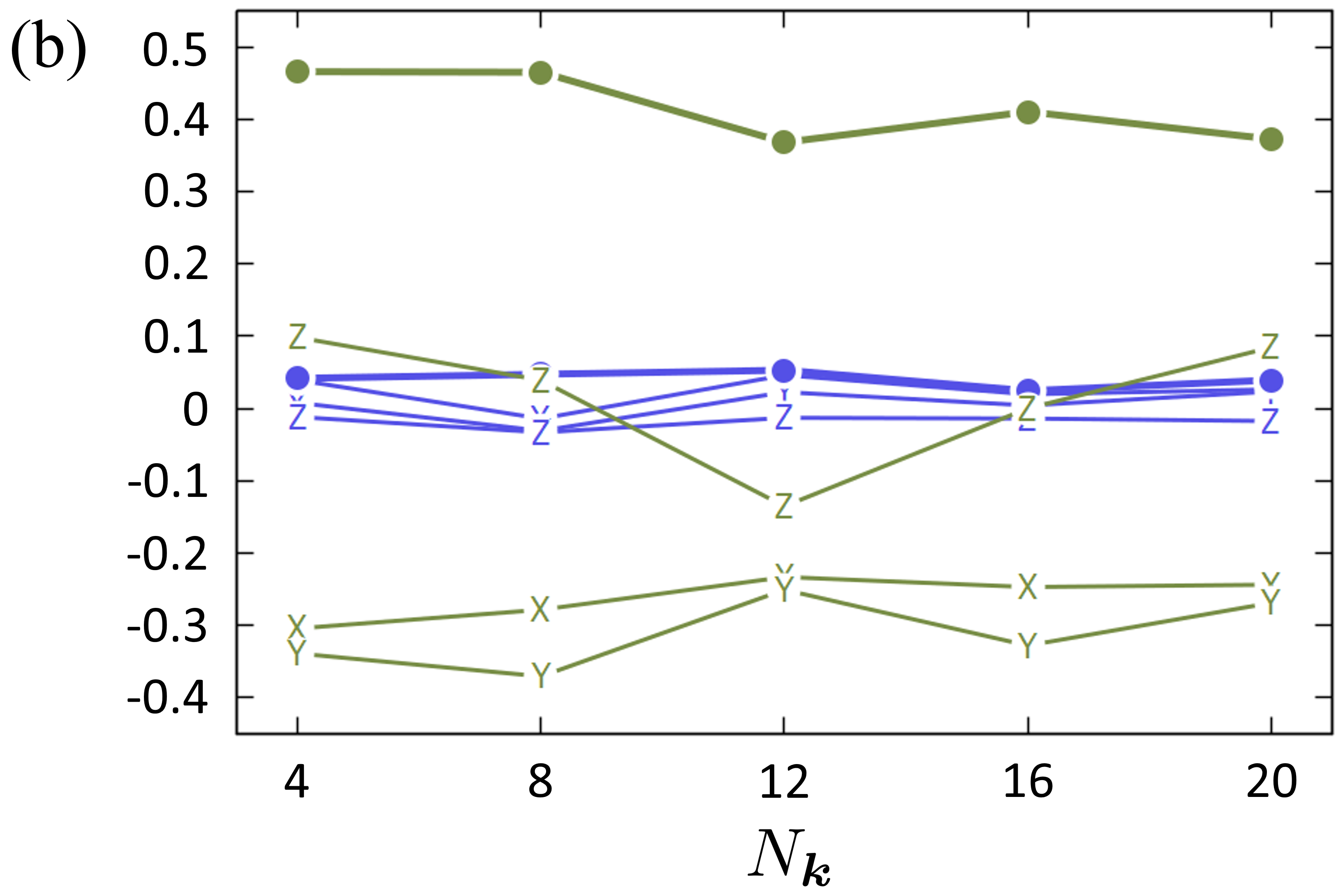}
\end{center}

\caption{\label{fig:FGTDMI} 
Obtained DMI vectors and constants $|\bm{D}_{ij}|$ in \FGT monolayer, in meV.
Labels 'X', 'Y', 'Z' and dots denote 
$D^{x}_{ij}$, 
$D^{y}_{ij}$,
$D^{z}_{ij}$ and
$|\bm{D}_{ij}|$, correspondingly.
Green / purple color describe the nearest neighbor atom couple, where $i$ atom is from the top layer and $j$ atom is from the middle / bottom layer 
(figure \ref{fig:FGTCrystalStructure}, local atomic positions are presented in \ref{AppendixA}).
(a)~Convergence dynamics of \EqRef{Eq:DMIinitial} as the maximum distance threshold ($R_{max}$, in \AA) of the sum $\sum_{p}$ grows. 
The arrowed values are the estimation using suggested approach \EqRef{Eq:DMIfinal}.
(b)~Estimations performed using the suggested approach \EqRef{Eq:DMIfinal} with Monkhorst-Pack grid density $N_{\bm{k}} \times N_{\bm{k}}$ (the arrowed values in (a) are those for the case of $20 \times 20$ grid).
}
\end{figure}

In figure~\ref{fig:FGTDMI} 
we present the estimated DMI vectors, obtained for the nearest neighbors of the Fe atoms constituting the top layer of the material 
(figure~\ref{fig:FGTCrystalStructure}). 
One can readily see that apart from general stability by means Monkhorst-Pack grid density, 
a similar convergence tendency takes place 
with respect to the threshold distance of the spatial sum $\sum_{p}$ in 
\EqRef{Eq:DMIinitial}, 
which acquires the reliable convergence criterion in the value found using the suggested method 
\EqRef{Eq:DMIfinal}.
As it was established in 
\cite{PhysRevB.102.060402} DMI is found essential only in interplay between top (or bottom) Fe layer and the middle one, whereas the top-bottom layers coupling is suppressed due to the mirror symmetry.
it is also important to note that DMI constants $|\bm{D}_{ij}|$ are in a reasonable agreement with the previous theoretical studies of the Fe$_3$GeTe$_2$-based materials, where the Green's functions apparatus was also employed 
\cite{PhysRevB.107.104428}. 

\section{Conclusion}

In the present work we have achieved a solution to the problem encountered by researchers when they consider microscopic anisotropic effects in real conducting materials and use the Green's function apparatus.
It was shown that the explicit use of $\bm{k}$-dependent Green's functions instead of inter-atomic ones makes it possible to carry out estimates of magnetic torque vector and Dzyaloshinskii-Moriya interactions with guaranteed numerical stability of spatial sums over the crystal lattice.
Furthermore, the time required for the computer calculation can be reduced up to two orders of magnitude, which for the first time makes it technically accessible to study materials using a more dense Monkhorst-Pack grid to describe the first Brillouin zone.

The proposed approaches were tested on the case of \FGT monolayer. 
Based on the calculated magnetic torque vector values, the typical sizes of the collective spin excitations in the domain structure were obtained in agreement with the experimental data, and obtained DMI constants are in qualitative accord with previous theoretical studies.
In this regard, we can consider the approaches as having a significant prospect for further application to the study of anisotropic effects in real conducting materials, as well as in complex heterostructures, in which the first-principles calculations and further application of the Green's function apparatus are greatly hampered by excessive computational cost.






\section{Acknowledgments}

The work is supported by the
Russian Science Foundation, 
Grant No. 23-72-01003,
https://rscf.ru/project/23-72-01003/







\appendix
\section{\textit{Ab initio} calculations of \FGT monolayer}
\label{AppendixA}

Our study of \FGT monolayer is based on the first-principles calculations of electronic structure within density functional theory (DFT) 
\cite{PhysRev.140.A1133} 
using generalized gradient approximation.
As the exchange-correlation functional we employ Perdew-Burke-Ernzerhof (PBE) 
\cite{PhysRevLett.77.3865}.
To perform the numerical computations the \textit{Quantum-Espresso} simulation package is used
\cite{Giannozzi_2009}.

\begin{figure}[htbp]
\begin{center}
\includegraphics[width=0.75\textwidth]{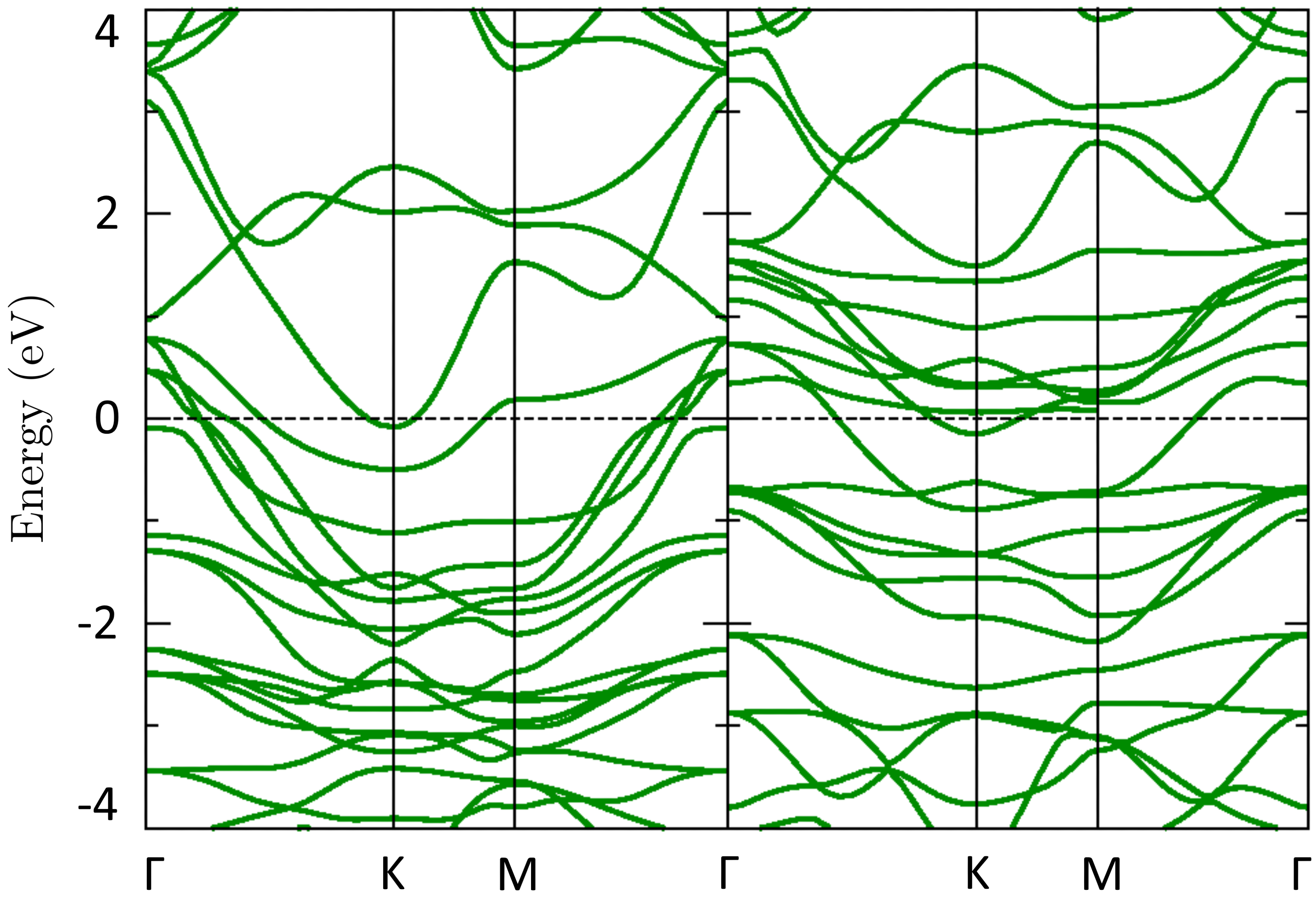}
\end{center}

\caption{\label{fig:FGTBandStructure} 
Resulting DFT band structure of \FGT monolayer 
(majority spin - \textit{left}, minority spin - \textit{right}). 
The high symmetry points are
 $\Gamma(0, 0, 0)$,
 $\mathrm{K}(0.333, 0.333, 0)$,
 $\mathrm{M}(0, 0.5, 0)$.
 The Fermi level is zero.
}
\end{figure}

The basic parameters of the simulation are following:
\begin{itemize}
\item The energy cutoff of the plane wave basis construction is set
to 330~eV;
\item The energy convergence criterion is $10^{-6}$~eV;
\item The $20 \times 20 \times 1$ Monkhorst-Pack grid was employed to carry out integration over the Brillouin zone;
\item The lattice vectors are:
$\bm{a}~=~( 3.991,     0)$~\AA, 
$\bm{b}~=~(-1.996, 3.456)$~\AA;
\item The local positions of three Fe atoms in the unit cell are:
$\bm{r}_1~=~(0,      0,       0)$~\AA,
$\bm{r}_2~=~(0,  2.304,   1.277)$~\AA,
$\bm{r}_3~=~(0,      0,   2.554)$~\AA.
\end{itemize}

Figure~\ref{fig:FGTBandStructure} gives the resulting band structure, in consistent with the previous theoretical investigations 
\cite{Badrtdinov}. 
In order to parametrize the low-energy model of magnetoactive Fe $3d$-shell we perform a projection resulting wave functions onto maximally localized Wannier functions
~\cite{PhysRevB.56.12847, PhysRevB.65.035109} 
using \textit{Wannier90} software 
\cite{MOSTOFI2008685}.
It contains with the Fe($3d$) states  
also $s,~p$ states of Fe, Ge and Te, 
due to the essential entanglement of corresponding bands.






\section*{References}

\bibliographystyle{iopart-num-long}

\bibliography{Biblio}

\providecommand{\newblock}{}
\begin{thebibliography}{10}
\expandafter\ifx\csname url\endcsname\relax
  \def\url#1{{\tt #1}}\fi
\expandafter\ifx\csname urlprefix\endcsname\relax\def\urlprefix{URL }\fi
\providecommand{\eprint}[2][arXiv]{#1:\linebreak[0]#2}

\bibitem{10.1063/5.0072735}
Marrows C~H and Zeissler K 2021 Perspective on skyrmion spintronics {\em Appl. Phys. Lett.\/} {\bf 119} 250502 ISSN 0003-6951 \urlprefix\url{https://doi.org/10.1063/5.0072735}

\bibitem{10.1126/science.1240573}
Romming N, Hanneken C, Menzel M, Bickel J~E, Wolter B, von Bergmann K, Kubetzka A and Wiesendanger R 2013 Writing and deleting single magnetic skyrmions {\em Science\/} {\bf 341} 636--639 (\textit{Preprint} \eprint{https://www.science.org/doi/pdf/10.1126/science.1240573}) \urlprefix\url{https://www.science.org/doi/abs/10.1126/science.1240573}

\bibitem{10.1063/5.0046950}
Vakili H, Xu J~W, Zhou W, Sakib M~N, Morshed M~G, Hartnett T, Quessab Y, Litzius K, Ma C~T, Ganguly S, Stan M~R, Balachandran P~V, Beach G~S~D, Poon S~J, Kent A~D and Ghosh A~W 2021 {Skyrmionics—Computing} and memory technologies based on topological excitations in magnets {\em J. Appl. Phys.\/} {\bf 130} 070908 ISSN 0021-8979 (\textit{Preprint} \eprint{https://pubs.aip.org/aip/jap/article-pdf/doi/10.1063/5.0046950/20021885/070908\_1\_5.0046950.pdf}) \urlprefix\url{https://doi.org/10.1063/5.0046950}

\bibitem{Ornelas2025}
Ornelas P, Nape I, de~Mello~Koch R and Forbes A 2025 Topological rejection of noise by quantum skyrmions {\em Nat. Commun.\/} {\bf 16} 2934 ISSN 2041-1723 \urlprefix\url{https://doi.org/10.1038/s41467-025-58232-4}

\bibitem{Huang_2017}
Huang B, Clark G, Navarro-Moratalla E, Klein D, Cheng R, Seyler K, Zhong D, Schmidgall E, McGuire M, Cobden D, Yao W, Xiao D, Jarillo-Herrero P and Xu X 2017 Layer-dependent ferromagnetism in a {van der Waals} crystal down to the monolayer limit {\em Nature\/} {\bf 546}

\bibitem{Gong_2017}
Gong C, Li L, Li Z, Ji H, Stern A, Xia Y, Cao T, Bao W, Wang C, Wang Y, Qiu Z, Cava R, Louie S, Xia J and Zhang X 2017 Discovery of intrinsic ferromagnetism in two-dimensional {van der Waals} crystals {\em Nature\/} {\bf 546}

\bibitem{Jiang2018}
Jiang S, Shan J and Mak K~F 2018 Electric-field switching of two-dimensional {van der Waals} magnets {\em Nat. Mater.\/} {\bf 17} 406--410 ISSN 1476-4660 \urlprefix\url{https://doi.org/10.1038/s41563-018-0040-6}

\bibitem{Huang_2018}
Huang B, Clark G, Klein D, MacNeill D, Navarro-Moratalla E, Seyler K, Wilson N, McGuire M, Cobden D, Xiao D, Yao W, Jarillo-Herrero P and Xu X 2018 Electrical control of 2d magnetism in bilayer {CrI}$_3$ {\em Nat. Nanotechnol.\/} {\bf 13}

\bibitem{PhysRevB.98.144411}
Webster L and Yan J~A 2018 Strain-tunable magnetic anisotropy in monolayer {CrCl}$_{3}$, {CrBr}$_{3}$, and {CrI}$_{3}$ {\em Phys. Rev. B\/} {\bf 98}(14) 144411 \urlprefix\url{https://link.aps.org/doi/10.1103/PhysRevB.98.144411}

\bibitem{adma.201900065}
Li H, Ruan S and Zeng Y~J 2019 Intrinsic van der {Waals} magnetic materials from bulk to the {2D} limit: {New} frontiers of spintronics {\em Adv. Mater.\/} {\bf 31} 1900065 (\textit{Preprint} \eprint{https://advanced.onlinelibrary.wiley.com/doi/pdf/10.1002/adma.201900065}) \urlprefix\url{https://advanced.onlinelibrary.wiley.com/doi/abs/10.1002/adma.201900065}

\bibitem{Deiseroth2006Fe3GeTe2AN}
Deiseroth H~J, Aleksandrov K, Reiner C, Kienle L and Kremer R~K 2006 {Fe}$_3${GeTe}$_2$ and {Ni}$_3${GeTe}$_2$ – {Two} new layered transition‐metal compounds: {Crystal} structures, {HRTEM} investigations, and magnetic and electrical properties {\em Eur. J. Inorg. Chem.\/} {\bf 2006} 1561--1567 \urlprefix\url{https://api.semanticscholar.org/CorpusID:97260784}

\bibitem{Deng_2018}
Deng Y, Yu Y, Song Y, Zhang J, Wang N~Z, Sun Z, Yi Y, Wu Y~Z, Wu S, Zhu J, Wang J, Chen X~H and Zhang Y 2018 Gate-tunable room-temperature ferromagnetism in two-dimensional {Fe}$_3${GeTe}$_2$ {\em Eur. J. Inorg. Chem.\/} {\bf 563} 94--99 \urlprefix\url{https://doi.org/10.1038/s41586-018-0626-9}

\bibitem{Chen_2013}
Chen B, Jinhu Y, Wang H, Imai M, Ohta H, Michioka C, Yoshimura K and Fang M 2013 Magnetic properties of layered itinerant electron ferromagnet {Fe}$_3${GeTe}$_2$ {\em J. Phys. Soc. Jpn.\/} {\bf 82} 124711

\bibitem{Fei}
Fei Z, Huang B, Malinowski P, Wang W, Song T, Sanchez J, Yao W, Xiao D, Zhu X, May A, Wu W, Cobden D, Chu J~H and Xu X 2018 Two-dimensional itinerant ising ferromagnetism in atomically thin {Fe}$_3${GeTe}$_2$ {\em Nat. Mater.\/} {\bf 17}

\bibitem{Kim_2018}
Kim K, Seo J, Lee E, Ko K~T, Kim B, Jang B~G, Ok J~M, Lee J, Jo Y, Kang W, Shim J, Kim C, Yeom H, Min B, Yang B~J and Kim J~S 2018 Large anomalous hall current induced by topological nodal lines in a ferromagnetic van der {Waals} semimetal {\em Nat. Mater.\/} {\bf 17} 794

\bibitem{Roemer_2020}
Roemer R, Liu C and Zou K 2020 Robust ferromagnetism in wafer-scale monolayer and multilayer {Fe}$_3${GeTe}$_2$ {\em NPJ 2D Mater. Appl.\/} {\bf 4} 33

\bibitem{PhysRevB.102.060402}
Laref S, Kim K~W and Manchon A 2020 Elusive {Dzyaloshinskii-Moriya} interaction in monolayer {Fe}$_3${GeTe}$_2$ {\em Phys. Rev. B\/} {\bf 102}(6) 060402 \urlprefix\url{https://link.aps.org/doi/10.1103/PhysRevB.102.060402}

\bibitem{acs.nanolett.9b03453}
Ding B, Li Z, Xu G, Li H, Hou Z, Liu E, Xi X, Xu F, Yao Y and Wang W 2020 Observation of magnetic skyrmion bubbles in a van der {Waals} ferromagnet {Fe}$_3${GeTe}$_2$ {\em Nano Lett.\/} {\bf 20} 868--873 (\textit{Preprint} \eprint{https://doi.org/10.1021/acs.nanolett.9b03453}) \urlprefix\url{https://doi.org/10.1021/acs.nanolett.9b03453}

\bibitem{PhysRevB.103.104410}
Park T~E, Peng L, Liang J, Hallal A, Yasin F~S, Zhang X, Song K~M, Kim S~J, Kim K, Weigand M, Sch\"utz G, Finizio S, Raabe J, Garcia K, Xia J, Zhou Y, Ezawa M, Liu X, Chang J, Koo H~C, Kim Y~D, Chshiev M, Fert A, Yang H, Yu X and Woo S 2021 N\'eel-type skyrmions and their current-induced motion in van der {Waals} ferromagnet-based heterostructures {\em Phys. Rev. B\/} {\bf 103}(10) 104410 \urlprefix\url{https://link.aps.org/doi/10.1103/PhysRevB.103.104410}

\bibitem{Birch}
Birch M~T, Powalla L, Wintz S, Hovorka O, Litzius K, Loudon J~C, Turnbull L~A, Nehruji V, Son K, Bubeck C, Rauch T~G, Weigand M, Goering E, Burghard M and Schütz G 2022 History-dependent domain and skyrmion formation in {2D} van der {Waals} magnet {Fe}$_3${GeTe}$_2$ {\em Nat. Commun.\/} {\bf 13} 3035 \urlprefix\url{https://www.nature.com/articles/s41467-022-30740-7}

\bibitem{wang2019directobservationschiralspin}
Wang H, Wang C, Zhu Y, Li Z~A, Zhang H, Tian H, Shi Y, Yang H and Li J 2019 Direct observations of chiral spin textures in van der {Waals} magnet {Fe}$_3${GeTe}$_2$ nanolayers (\textit{Preprint} \eprint{1907.08382}) \urlprefix\url{https://arxiv.org/abs/1907.08382}

\bibitem{PhysRevB.71.184434}
Mazurenko V~V and Anisimov V~I 2005 Weak ferromagnetism in antiferromagnets: $\alpha$-{Fe}$_{2}${O}$_{3}$ and {La}$_{2}${Cu}{O}$_{4}$ {\em Phys. Rev. B\/} {\bf 71}(18) 184434 \urlprefix\url{https://link.aps.org/doi/10.1103/PhysRevB.71.184434}

\bibitem{PhysRevB.89.214422}
Mazurenko V~V, Kvashnin Y~O, Jin F, De~Raedt H~A, Lichtenstein A~I and Katsnelson M~I 2014 First-principles modeling of magnetic excitations in {Mn}$_{12}$ {\em Phys. Rev. B\/} {\bf 89}(21) 214422 \urlprefix\url{https://link.aps.org/doi/10.1103/PhysRevB.89.214422}

\bibitem{PhysRevB.64.174402}
Pajda M, Kudrnovsk\'y J, Turek I, Drchal V and Bruno P 2001 Ab initio calculations of exchange interactions, spin-wave stiffness constants, and curie temperatures of {Fe}, {Co}, and {Ni} {\em Phys. Rev. B\/} {\bf 64}(17) 174402 \urlprefix\url{https://link.aps.org/doi/10.1103/PhysRevB.64.174402}

\bibitem{KASHIN201858}
Kashin I~V, Andreev S~N and Mazurenko V~V 2018 First-principles study of isotropic exchange interactions and spin stiffness in {FeGe} {\em J. Magn. Magn. Mater.\/} {\bf 467} 58--63 ISSN 0304-8853 \urlprefix\url{https://www.sciencedirect.com/science/article/pii/S0304885318308916}

\bibitem{PhysRevB.106.134434}
Kashin I~V, Gerasimov A and Mazurenko V~V 2022 Reciprocal space study of {Heisenberg} exchange interactions in ferromagnetic metals {\em Phys. Rev. B\/} {\bf 106}(13) 134434 \urlprefix\url{https://link.aps.org/doi/10.1103/PhysRevB.106.134434}

\bibitem{kashin2024magnetocrystalline}
Kashin I~V and Andreev S~N 2024 Magnetocrystalline anisotropy in metallic systems: fast and stable estimation in {Green}'s functions formalism {\em arXiv preprint arXiv:2403.14241\/}

\bibitem{PhysRevB.13.5188}
Monkhorst H~J and Pack J~D 1976 Special points for {Brillouin-zone} integrations {\em Phys. Rev. B\/} {\bf 13}(12) 5188--5192 \urlprefix\url{https://link.aps.org/doi/10.1103/PhysRevB.13.5188}

\bibitem{PhysRevB.52.13419}
Solovyev I~V, Dederichs P~H and Mertig I 1995 Origin of orbital magnetization and magnetocrystalline anisotropy in {TX} ordered alloys (where {T=Fe,Co} and {X=Pd,Pt}) {\em Phys. Rev. B\/} {\bf 52}(18) 13419--13428 \urlprefix\url{https://link.aps.org/doi/10.1103/PhysRevB.52.13419}

\bibitem{PhysRevB.39.865}
Bruno P 1989 Tight-binding approach to the orbital magnetic moment and magnetocrystalline anisotropy of transition-metal monolayers {\em Phys. Rev. B\/} {\bf 39}(1) 865--868 \urlprefix\url{https://link.aps.org/doi/10.1103/PhysRevB.39.865}

\bibitem{PhysRevB.47.14932}
Wang D~s, Wu R and Freeman A~J 1993 First-principles theory of surface magnetocrystalline anisotropy and the diatomic-pair model {\em Phys. Rev. B\/} {\bf 47}(22) 14932--14947 \urlprefix\url{https://link.aps.org/doi/10.1103/PhysRevB.47.14932}

\bibitem{Goringe_1997}
Goringe C~M, Bowler D~R and Hernández E 1997 Tight-binding modelling of materials {\em Rep. Prog. Phys.\/} {\bf 60} 1447 \urlprefix\url{https://dx.doi.org/10.1088/0034-4885/60/12/001}

\bibitem{LocalForceTheorem_1}
Machintosh A and Andersen O 1980 {\em Electrons at the {Fermi} surface\/} (Cambridge University Press) p 149

\bibitem{LocalForceTheorem_2}
Methfessel M and Kubler J 1982 Bond analysis of heats of formation: application to some group {VIII} and {IB} hydrides {\em J. Phys. F: Met. Phys.\/} {\bf 12} 141--161 \urlprefix\url{https://doi.org/10.1088/0305-4608/12/1/013}

\bibitem{Lichtenstein2013correl13}
Lichtenstein A 2013 {\em Magnetism: {From} {S}toner to {H}ubbard\/} ({Forschungszentrum Jülich GmbH Institute for Advanced Simulations, Jülich, Germany}) \urlprefix\url{https://www.cond-mat.de/events/correl13/manuscripts/lichtenstein.pdf}

\bibitem{PhysRevLett.77.3865}
Perdew J~P, Burke K and Ernzerhof M 1996 Generalized gradient approximation made simple {\em Phys. Rev. Lett.\/} {\bf 77}(18) 3865--3868 \urlprefix\url{https://link.aps.org/doi/10.1103/PhysRevLett.77.3865}

\bibitem{PhysRevB.56.12847}
Marzari N and Vanderbilt D 1997 {Maximally localized generalized Wannier functions for composite energy bands} {\em Phys. Rev. B\/} {\bf 56}(20) 12847--12865 \urlprefix\url{https://link.aps.org/doi/10.1103/PhysRevB.56.12847}

\bibitem{PhysRevB.65.035109}
Souza I, Marzari N and Vanderbilt D 2001 Maximally localized wannier functions for entangled energy bands {\em Phys. Rev. B\/} {\bf 65}(3) 035109 \urlprefix\url{https://link.aps.org/doi/10.1103/PhysRevB.65.035109}

\bibitem{Deiseroth}
Deiseroth H~J, Aleksandrov K, Reiner C, Kienle L and Kremer R~K 2006 {Fe}$_3${GeTe}$_2$ and {Ni}$_3${GeTe}$_2$ – {Two} new layered transition-metal compounds: {Crystal} structures, {HRTEM} investigations, and magnetic and electrical properties {\em Eur. J. Inorg. Chem.\/} {\bf 2006} 1561--1567 \urlprefix\url{https://chemistry-europe.onlinelibrary.wiley.com/doi/abs/10.1002/ejic.200501020}

\bibitem{PUSHKAREV2023171456}
Pushkarev G~V, Badrtdinov D~I, Iakovlev I~A, Mazurenko V~V and Rudenko A~N 2023 An effective spin model on the honeycomb lattice for the description of magnetic properties in two-dimensional {Fe}$_3${GeTe}$_2$ {\em J. Magn. Magn. Mater.\/} {\bf 588} 171456 ISSN 0304-8853 \urlprefix\url{https://www.sciencedirect.com/science/article/pii/S030488532301106X}

\bibitem{LIECHTENSTEIN198765}
Liechtenstein A, Katsnelson M, Antropov V and Gubanov V 1987 Local spin density functional approach to the theory of exchange interactions in ferromagnetic metals and alloys {\em J. Magn. Magn. Mater.\/} {\bf 67} 65--74 ISSN 0304-8853 \urlprefix\url{https://www.sciencedirect.com/science/article/pii/0304885387907219}

\bibitem{Badrtdinov}
Badrtdinov D, Pushkarev G, Katsnelson M and Rudenko A 2023 Electron transport and scattering mechanisms in ferromagnetic monolayer {Fe}$_3${GeTe}$_2$ {\em npj 2D Mater. Appl.\/} {\bf 7} 52 \urlprefix\url{https://www.nature.com/articles/s41699-023-00413-0}

\bibitem{Deng}
Deng Y, Yu Y, Song Y, Zhang J, Wang N~Z, Sun Z, Yi Y, Wu Y~Z, Wu S, Zhu J, Wang J, Chen X~H and Zhang Y 2018 Gate-tunable room-temperature ferromagnetism in two-dimensional {Fe}$_3${GeTe}$_2$ {\em Nature\/} {\bf 563} 94 \urlprefix\url{https://www.nature.com/articles/s41586-018-0626-9}

\bibitem{Zaiyao}
Fei Z, Huang B, Malinowski P, Wang W, Song T, Sanchez J, Yao W, Xiao D, Zhu X, May A~F, Wu W, Cobden D~H, Chu J~H and Xu X 2018 Two-dimensional itinerant ferromagnetism in atomically thin {Fe}$_3${GeTe}$_2$ {\em Nature Mater.\/} {\bf 17} 778 \urlprefix\url{https://www.nature.com/articles/s41563-018-0149-7}

\bibitem{PhysRevB.107.104428}
Li D, Haldar S, Drevelow T and Heinze S 2023 Tuning the magnetic interactions in van der waals {Fe}$_3${GeTe}$_2$ heterostructures: {A} comparative study of ab initio methods {\em Phys. Rev. B\/} {\bf 107}(10) 104428 \urlprefix\url{https://link.aps.org/doi/10.1103/PhysRevB.107.104428}

\bibitem{PhysRev.140.A1133}
Kohn W and Sham L~J 1965 Self-consistent equations including exchange and correlation effects {\em Phys. Rev.\/} {\bf 140}(4A) A1133--A1138 \urlprefix\url{https://link.aps.org/doi/10.1103/PhysRev.140.A1133}

\bibitem{Giannozzi_2009}
Giannozzi P, Baroni S, Bonini N, Calandra M, Car R, Cavazzoni C, Ceresoli D, Chiarotti G~L, Cococcioni M, Dabo I, Corso A~D, de~Gironcoli S, Fabris S, Fratesi G, Gebauer R, Gerstmann U, Gougoussis C, Kokalj A, Lazzeri M, Martin-Samos L, Marzari N, Mauri F, Mazzarello R, Paolini S, Pasquarello A, Paulatto L, Sbraccia C, Scandolo S, Sclauzero G, Seitsonen A~P, Smogunov A, Umari P and Wentzcovitch R~M 2009 {QUANTUM ESPRESSO: a modular and open-source software project for quantum simulations of materials} {\em J. Phys.: Condens. Matter\/} {\bf 21} 395502 \urlprefix\url{https://dx.doi.org/10.1088/0953-8984/21/39/395502}

\bibitem{MOSTOFI2008685}
Mostofi A~A, Yates J~R, Lee Y~S, Souza I, Vanderbilt D and Marzari N 2008 {wannier90: A tool for obtaining maximally-localised Wannier functions} {\em Comput. Phys. Commun.\/} {\bf 178} 685--699 ISSN 0010-4655 \urlprefix\url{https://www.sciencedirect.com/science/article/pii/S0010465507004936}

\end{thebibliography}



\end{document}